\documentstyle[epsbox,12pt]{article}

\begin{document}

\newcommand{\dfrac}[2]{\displaystyle{\frac{#1}{#2}}}

{\it University of Shizuoka}

\hspace*{11.5cm} {\bf US-98-05}\\[-.3in]

\hspace*{11.5cm} {\bf May 1998}\\[.3in]

%\vspace*{.1in}

\begin{center}
{\large\bf  Universal Seesaw Mass Matrix Model}\\[.2in]

{\large\bf  and Neutrino Phenomenology }\footnote{
Contributed paper to XVIII International Conference on
{\it Neutrion Physics and Astrophysics} (presented
at the poster session), Takayama, 
Japan, June 4-9, 1998.}\\[.3in]

{\bf Yoshio Koide}\footnote{
E-mail: koide@u-shizuoka-ken.ac.jp} \\

Department of Physics, University of Shizuoka \\ 
52-1 Yada, Shizuoka 422-8526, Japan \\[.1in]

\vspace{.3in}

{\large\bf Abstract}
\end{center}

\begin{quotation}
Stimulated by the recent development of the 
``universal seesaw mass matrix model", an application 
of the model to the neutrino mass matrix is investigated:
For the charged lepton and down-quark sectors, the 
model explains the smallness of their masses $m_f$ by 
the conventional seesaw mechanism  $M_f\simeq m_L M_F^{-1}m_R$
($M_F$ is a mass matrix of hypothetical heavy fermions $F$).
On the other hand, the observed fact 
$m_t\sim \Lambda_L=O(m_L)$
(electroweak scale $\Lambda_L=174$ GeV) 
seems to reject the applying of the 
seesaw mechanism to the up-quark sector. 
However, recently, it has been found that, 
by taking det$M_F=0$ for the up-quark sector $F=U$, 
we can understand the question of why only top quark
has a mass of the order of $\Lambda_L$ without  
the sesaw-suppression factor $O(m_R)/O(M_F)$. 
For neutrino sector, the mass matrix $M_\nu$ is given by
$M_\nu \simeq m_L M_F^{-1} m_L^T$ ($F=N$), so that the masses 
$m_\nu$ are suppressed by a factor $O(m_L)/O(m_R)$ 
compared with the conventional quark and charged lepton
masses.
The model can naturally lead to a large mixing 
$\sin^2 2\theta \simeq 1$.
Also another model is investigated within the framework of 
the universal seesaw model: the model leads to three sets
of the almost degenerate two Majorana neutrinos which
are large mixing states between the left-handed neutrinos
$\nu_{Li}$ and SU(2)$_L\times$SU(2)$_R$ singlet neutrinos
$N_{1i}$ ($i=e,\mu,\tau$), so that the model can give a 
simultaneous explanation of the atmospheric and solar
neutrino data.
\end{quotation}

%%%%%%%%%%%%%%%%%%%%%%%%%%%%%%%%%
\newpage
%%%%%%%%%%%%%%%

{\large\bf 1. Introduction}

\vglue.05in

Recent progress of the non-accelerator and accelerator 
experiments has gloriously accumulated knowledge of the 
masses and mixings of neutrinos.
One of  the most challenging problems in the particle 
physics is to give a unified understanding of quark and lepton
masses and mixings.
Especially, the study of the neutrino mass matrix will give us 
a valuable clue to the unified understanding of the quarks 
and leptons.
Now, the study of the unified mass matrix models is just timely.

As one of such the unified mass matrix models, the so-called 
``universal seesaw" mass matrix model [1] has recently revived.
The seesaw mechanism  was first proposed [2] in order to answer the question 
of why neutrino masses are so invisibly small.
Then, in order to understand that the observed quark and lepton 
masses are considerably smaller than the electroweak scale,
the mechanism was applied to the quarks [1]:
A would-be seesaw mass matrix for $(f, F)$ is
expressed as
$$
M = \left(\begin{array}{cc}
0 & m_L \\
m_R & M_F \\ 
\end{array} \right) = m_0 \left( 
\begin{array}{cc}
0 & Z_L \\
\kappa Z_R & \lambda Y_f \\
\end{array} \right) \ \ , 
\eqno(1.1)
$$
where $f=u, d, \nu, e$ are the conventional quarks and leptons,
$F=U, D, N, E$ are hypothetical heavy fermions, and 
they belong to $f_L = (2,1)$, $f_R = (1,2)$, 
$F_L = (1,1)$ and $F_R = (1,1)$ of 
SU(2)$_L \times $SU(2)$_R$.
The matrices $Z_L$, $Z_R$ and $Y_f$ are of the order one.
The matrices $m_L$ and $m_R$ take universal structures for
quarks and leptons.
Only the heavy fermion matrix $M_F$ takes a structure 
dependent on $f=u,d,\nu,e$.
For the case $\lambda \gg \kappa \gg 1$, 
the mass matrix (1.1) leads to the well-known seesaw expression
$$
 M_f \simeq -m_L M_F^{-1} m_R  \ .
\eqno(1.2)
$$

However, the observation of the top quark of 1994 [3] aroused
a question: Can the observed fact
$m_t \simeq 180$ GeV $\sim \Lambda_L = O(m_L)$ 
be accommodated to the universal seesaw mass matrix scenario?
Because $m_t\sim O(m_L)$ means $M_F^{-1}m_R \sim O(1)$.
For this question, a recent study gives the answer ``Yes":
Yes, we can do [4,5] by putting an additional constraint 
$$
{\rm det}M_F =0 \ .
\eqno(1.3)
$$
on the up-quark sector ($F=U$).
Then, we will easily be able to understand why only top quark $t$ 
acquires  the mass $m_t \sim O(m_L)$.
In the next section, we will review the mass generation scenario
on the basis of the universal seesaw mass matrix model with 
the constraint (1.3).
Also, in the next section, a rough sketch of the neutrino mass 
generation scenarios within the framework of the universal seesaw 
mass matrix model is given.
One (Model A) is a straightforward extension of (1.2), so that
the neutrino mass matrix $M_\nu$ is given by
$M_\nu \simeq - m_L M_N^{-1} m_R$.
The smallness of the neutrino masses $m_\nu \ll m_f$
($f=e,u,d$) is explained by assuming $\lambda_\nu
\gg \lambda$ ($\lambda\equiv \lambda_e=\lambda_u=\lambda_d$),
although the assumption is somewhat artificial.
Another one (Model B) is a model without a new scale 
parameter such as $\lambda_\nu$:
the light neutrino mass matrix $M_\nu$ is given by
$$
M_\nu \simeq - m_L M_N^{-1} m_L^T \ , 
\eqno(1.4)
$$
so that the light neutrino masses $m_\nu$ are given with the order of
$$
m_\nu \sim O\left(\frac{1}{\lambda} m_0\right) 
\sim O\left(\frac{1}{\kappa} m_{charged\ lepton}\right) \ .
\eqno(1.5)
$$
Therefore, the model B can explain the smallness of the 
neutrino masses without assuming an additional scale
parameter such as $\lambda_\nu$ in the model A.
The third scenario (Model C) is very attractive to the 
neutrino phenomenology, because the model can lead to 
three sets of almost degenerate two Majorana neutrinos
(the pseudo-Dirac neutrinos [6]). 
Every model of these can naturally give a large mixing 
$\sin^2 2\theta \simeq 1$.

In Sec.~3, we give a more explicit model of the universal
seesaw mass matrix with some special structures of 
$m_L$, $m_R$ and $M_F$.
In Secs.~4, 5, and 6, the models A, B, and C are discussed,
respectively.
Finally, Sec.~7 is devoted to the summary and concluding 
remarks.

\vspace*{.2in}
%%%%%%%%%%%%%%%

{\large\bf 2. Energy scales and fermion masses }

\vglue.05in

For convenience, we take the diagonal basis of the matrix $M_F$.
Then, the condition (1.3) means that the heavy fermion mass matrix
$M_F$ in the up-quark sector is given by
$$ 
M_U=\lambda m_0 \left(
\begin{array}{ccc}
O(1) & 0 & 0 \\
0 & O(1) & 0 \\
0 & 0 & 0 
\end{array} \right) \ ,
\eqno(2.1)
$$
although the other heavy fermion mass matrices $M_F$ ($F\neq U$) 
are given by
$$
M_F=\lambda m_0 \left(
\begin{array}{ccc}
O(1) & 0 & 0 \\
0 & O(1) & 0 \\
0 & 0 & O(1) 
\end{array} \right) \ , \ \  (F\neq U) \ . 
\eqno(2.2)
$$ 
Note that for the third up-quark the seesaw mechanism does
not work (see Fig.~1).

%%%%%%%%%%%%%%%%%%%%%%%%%%%%%%%%%%%%%%%%%%%
\vspace{.3in}
\begin{figure}[htb]

det$M_F\neq 0$ \ $\Longrightarrow$ \ Seesaw Mass
\hspace{17mm}
det$M_F = 0$ \ $\Longrightarrow$ \ Non-Seesaw Mass
\begin{picture}(650,160)(0,0)
\put(10,00){\thicklines\line(1,0){180}}
\put(100,00){\thicklines\line(0,1){120}}
\put(40,21){\thinlines\line(2,1){120}}
\put(40,19){\thinlines\line(2,1){120}}
\put(40,71){\thinlines\line(2,1){120}}
\put(40,69){\thinlines\line(2,1){120}}

\put(100,50){\circle*{5}}
\put(100,100){\circle*{5}}
\put(45,35){\circle{20}}
\put(40,32){$F_L$}
\put(40,82){{$F_R$}}
\put(45,85){\circle{20}}
\put(143,82){{$f_R$}}
\put(148,86){\circle{20}}
\put(143,132){{$f_L$}}
\put(148,136){\circle{20}}
\put(25,35){\thicklines\line(0,1){50}}
\put(25,35){\thicklines\line(1,0){8}}
\put(25,85){\thicklines\line(1,0){8}}
\put(5,53){{$M_F$}}
\put(105,35){{$\Lambda_R$}}
\put(105,85){{$\Lambda_L$}}
\put(105,5){{$\Lambda_S$}}
\put(145,10){{$m(F_L,F_R)\sim\Lambda_S$}}
\put(141,110){{$m(f_L,f_R)\sim\frac{\Lambda_L\Lambda_R}{\Lambda_S}$}}

%%%%%%%%%%%%%%%%%%%%%%%%%%%%%%%%%%%%%%%%%%%%%
\put(210,00){\thicklines\line(1,0){180}}
\put(300,00){\thicklines\line(0,1){120}}
\put(235,51){\thinlines\line(1,0){130}}
\put(235,49){\thinlines\line(1,0){130}}
\put(235,101){\thinlines\line(1,0){130}}
\put(235,99){\thinlines\line(1,0){130}}
\put(300,50){\circle*{5}}
\put(300,100){\circle*{5}}
\put(237,55){$F_{L3}$}
\put(245,59){\circle{21}}
\put(237,105){{$F_{R3}$}}
\put(245,109){\circle{21}}
\put(355,55){{$f_{R3}$}}
\put(362,59){\circle{21}}
\put(355,105){{$f_{L3}$}}
\put(362,109){\circle{21}}
\put(305,35){{$\Lambda_R$}}
\put(305,85){{$\Lambda_L$}}
\put(305,5){{$\Lambda_S$}}
\put(345,30){{$m(F_{L3},f_{R3})\sim\Lambda_R$}}
\put(345,80){{$m(f_{L3},R_{R3})\sim\Lambda_L$}}
\end{picture}

\vglue.1in

\centerline{Fig.~1. Seesaw and non-seesaw masses}
\end{figure}

\vspace{.2in}
%%%%%%%%%%%%%%%%%%%%%%%%%%%%%%%%%%%%%%%%%%%%%%%%%%%%%%%%%%%%%%%%

The mass generation at each energy scale is as follows.
First, at the energy scale $\mu=\Lambda_S$, the heavy fermions $F$,
except for $U_3$, acquire the masses of the order of $\Lambda_S$.
Second, at the energy scale $\mu=\Lambda_R$, the SU(2)$_R$ 
symmetry is broken, and the fermion $u_{R3}$ generates a mass
term of the order of $\Lambda_R$ by pairing with $U_{L3}$.
Finally, at $\mu=\Lambda_L$, the SU(2)$_L$ symmetry is broken,
and the fermion $u_{L3}$ generates a mass term of the order 
$\Lambda_L$ by pairing with $U_{R3}$. 
The other fermions $f$ acquire the well-known seesaw masses
(1.2).
The scenario is summarized in Table 1.

%%%%%%%%%%%%%%%%%%%%%%%%%%%%%%%%%%%%%%%%%%
\vspace{.2in}
\centerline{Table 1. Fermion mass generation scenario}
\vglue.05in
%\noindent
\begin{tabular}{|c|c|cc|}\hline
Energy scale &  $d$- \& $e$-sectors &
$u$-sector ($i=1,2$) & (i=3)\\ \hline
{\large At} $\mu=\Lambda_S\sim \lambda m_0$ & $m(F_L, F_R) \sim \Lambda_S$  
& $m(U_{Li}, U_{Ri}) \sim \Lambda_S$ &  $m(U_{L3}, U_{R3}) =0$ \\ \hline
{\large At} $\mu=\Lambda_R\sim \kappa m_0$ & $m(f_R, F_L)\sim \Lambda_R$ 
& $m(u_{Ri}, U_{Li}) \sim \Lambda_R$ & $m(u_{R3}, U_{L3}) \sim \Lambda_R$ 
 \\ \hline 
{\large At} $\mu=\Lambda_L\sim m_0$ & $m(f_L,F_R)\sim \Lambda_L$ & 
$m(u_{Li},U_{Ri})\sim \Lambda_L$ & $m(u_{L3}, U_{R3})\sim \Lambda_L$  \\[.2in] 
 &  $\Downarrow $ \ \ \ \ \  & $\Downarrow$ \ \ \ \ \  &  \\ 
& $m(f_L, f_R)\sim \displaystyle\frac{\Lambda_L \Lambda_R}{\Lambda_S}$  
& $m(u_{Li}, u_{Ri}) \sim \displaystyle\frac{\Lambda_L \Lambda_R}{\Lambda_S}$ 
& \\ \hline
\end{tabular}

\vglue.1in

Thus, we can understand why only top quark $t$ acquires  
the mass $m_t \sim O(m_L)$.
The other quarks and charged leptons acquire masses suppressed
by a factor $\kappa/\lambda=\Lambda_R/\Lambda_S$.
A suitable choice of the mass matrix parameters will be able
to give reasonable quark masses and Cabibbo-Kobayashi-Maskawa
[7] (CKM) matrix parameters (for example, see the next section).

Next, we discuss the neutrino mass generation.
Within the framework of the universal seesaw mass matrix model, 
we will discuss the following three scenarios.

One (Scenario A) is a trivial extension of 
the present model:
we introduce a further large energy scale 
$\Lambda_{S}^\nu$ in addition to $\Lambda_S$, and
we assume that $M_F \sim \Lambda_S$ ($F=U,D,E$), while 
$M_N \sim \Lambda_{S}^\nu$ ($\Lambda_{S}^\nu\gg \Lambda_S$).
The mass matrix $M_\nu$ for the conventional light 
neutrinos (Dirac neutrinos) is given by 
$$
M_\nu \simeq - m_L M_N^{-1} m_R \ .
\eqno(2.3)
$$
The Dirac neutrino masses $m_i^\nu$ are suppressed by a 
factor $\Lambda_S/\Lambda_S^\nu$ compared with the 
charged lepton masses $m_i^e$.

Another one (Scenario B) is more attractive 
because we does not introduce an additional energy scale.
The neutral heavy leptons are singlets of 
SU(2)$_L\times$SU(2)$_R$ and they do not have
U(1)-charge. 
Therefore, it is likely that they acquire 
Majorana masses $M_M$ together with the Dirac 
masses $M_D\equiv M_N$ at $\mu=\Lambda_S$.
Then, the mass matrix $M_\nu$ for the conventional light 
neutrinos (Majorana neutrinos) is given by
$$
M_\nu \simeq -m_L M_N^{-1} m_L^T \ , 
\eqno(2.4)
$$
so that the masses $m_i^\nu$ are given with the order of 
$$
m_\nu \sim \frac{\Lambda_L^2}{\Lambda_S}=
\frac{1}{\kappa}\frac{\Lambda_L\Lambda_R}{\Lambda_S}
\sim  O\left(\frac{1}{\kappa} m_i^e\right) \ ,
\eqno(2.5)
$$
where $\kappa=\Lambda_R/\Lambda_L$.
In order to explain the smallness of $m_\nu$, 
the model [8,9]
requires that the scale $\Lambda_R$ must be extremely 
larger than $\Lambda_L$.
The details are discussed in Sec.~5.

On the other hand, the scenario A allows a case with
a lower value of $\Lambda_R$.
If we consider $\kappa\sim 10$, then  we can expect abundant 
new physics effects as discussed in Ref.[10].
Therefore, although the model B is attractive from the 
theoretical point of view, the model A is also attractive
from the phenomenological point of view.
The model B is discussed in Sec.~4.

There is a model which is phenomenologically more 
attractive.
In the present model, there is no distinction between
$N_L$ and $N_R$, because both fields are 
SU(2)$_L\times$SU(2)$_R$ singles and do not have 
U(1) charges.
Therefore, if $\nu_L$ ($\nu_R$) acquire masses $m_L$ ($m_R$)
together with the partners $N_R$ ($N_L$), they may also 
acquire masses $m'_L$ ($m'_R$) together with the partners 
$N_L^c$ ($N_R^c$), where  $\psi^c$ denotes charge 
conjugate state $C\overline{\psi}^T$.
In the limit of $m'_L=m_L$ and $m'_R=m_R$, we obtain
six massless Majorana neutrino states: three are states
which consist of almost left-handed neutrino $\nu_L$, 
and the other are states which consist of 
$N_1 \equiv (N_L-N_R^c)/\sqrt{2}$.
A suitable choice of the differences between 
$m'_L$ and $m'_R$ and between $m'_R$ and $m_R$ can
lead to three sets of almost degenerate two Majorana 
neutrinos which are large mixing states between 
$\nu_L$ and $N_1$.
The almost degenerate states with a  large mixing between 
$\nu_{L\mu}$ and $N_{1\mu}$ and those with a large mixing
between $\nu_{L e}$ and $N_{1 e}$ are favorable to
the explanation of the observed data of the atmospheric
[11] and solar [12] neutrinos, respectively.
However, in the model C, we must introduce a new parameter
$\varepsilon$ which characterizes the differences
$m'_L-m_L$ and $m'_R-m_R$ by hand, differently from 
the model B.
The details of the model C are discussed in Sec.~6.
 
These  neutrino mass generation scenarios are 
roughly summarized in Table 2.

%%%%%%%%%%%%%%%%%%%%%%%%%%%%%%%%%
\vspace{.2in}

\centerline{Table 2. Neutrino mass generation scenarios}
\vglue.05in

\begin{tabular}{|c|c|c|c|}\hline
Energy scale &  Scenario A & Scenario B & Scenario C \\ \hline
{\large At} $\mu=\Lambda_{S}^\nu$ & $m(N_L, N_R) \sim \Lambda_{S}^\nu$  
&  & \\ \hline
{\large At} $\mu=\Lambda_S$ &  & $m(N_L, N_R) \sim \Lambda_S$  &
$m(N_L, N_R) \sim \Lambda_S$ \\ 
   &  & $m(N_L, N_L^c) \sim \Lambda_S$  & $m(N_L, N_L^c) \sim \Lambda_S$ \\ 
   &  & $m(N_R, N_R^c) \sim \Lambda_S$  & $m(N_R, N_R^c) \sim \Lambda_S$ \\ 
\hline
{\large At} $\mu=\Lambda_R$ & $m(\nu_R, N_L)\sim \Lambda_R$    
& $m(\nu_R, N_L)\sim \Lambda_R$ &
$m(\nu_R, N_L)\sim \Lambda_R$   \\ 
 & & & $m(\nu_R, N_R^c)\sim \Lambda_R$   \\  \hline 
{\large At} $\mu=\Lambda_L$ &  $m(\nu_L, N_R)\sim \Lambda_L$ 
& $m(\nu_L, N_R)\sim \Lambda_L$ & $m(\nu_L, N_R)\sim \Lambda_L$   \\
 & & & $m(\nu_L, N_L^c)\sim \Lambda_L$   \\  
 & $\Downarrow$ \ \ \ \ \ & $\Downarrow$ \ \ \ \ \ 
 & $\Downarrow$ \ \ \ \ \ \\ 
 & $m(\nu_L, \nu_R)\sim 
 \displaystyle\frac{\Lambda_L \Lambda_R}{\Lambda_{S}^\nu}$ &
$m(\nu_L, \nu_L^c) \sim \displaystyle\frac{\Lambda_L^2}{\Lambda_S}$
& $m(\nu_L,\nu_L^c)\simeq 0$  \\ \hline
\end{tabular}

\vspace{.2in}
%%%%%%%%%%%%%%%%%%%%%

\vspace{.2in}
%%%%%%%%%%%%%%%%%%%%%%%%%%%%%%%%%%%%%%%%%%%%%%%

{\large\bf 3. Democratic seesaw mass matrix model}

\vglue.05in

So far, we have not assumed explicit structures of the 
matrices $Z_L$, $Z_R$ and $Y_f$.
Here, in order to give a realistic numerical example,
we put the following working hypotheses [4]: 

\noindent
(i) The matrices $Z_L$ and $Z_R$, which are universal
for quarks and leptons, have the same 
structure:
$$
Z_L = Z_R \equiv Z = {\rm diag} (z_1, z_2, z_3) \ \ , 
\eqno(3.1)
$$
with $z_1^2 + z_2^2 + z_3^2 = 1$, 
where, for convenience, we have taken a basis on which 
the matrix $Z$ is diagonal. 

\noindent
(ii) The matrices $Y_f$, which have structures 
dependent on the fermion sector $f=u,d,\nu,e$, take
a simple form [(unit matrix)+(a rank one matrix)]:
$$
Y_f = {\bf 1} + 3 b_f X \ \ . 
\eqno(3.2)
$$
(iii) The rank one matrix is  given by
a democratic form
$$
X = \frac{1}{3}\left(\begin{array}{ccc}
1 & 1 & 1 \\
1 & 1 & 1 \\
1 & 1 & 1 \\
\end{array} \right) \  , 
\eqno(3.3)
$$
on the family-basis where the matrix $Z$ is diagonal.

\noindent
(iv) In order to fix the parameters $z_i$, we 
tentatively take $b_e = 0$ for the charged lepton sector,
so that the parameters $z_i$ are given by
$$
\frac{z_1}{\sqrt{m_e}} = \frac{z_2}{\sqrt{m_\mu}} = 
\frac{z_3}{\sqrt{m_\tau}} = \frac{1}{\sqrt{m_e + m_\mu + m_\tau}} \  , 
\eqno(3.4)
$$
from $M_e \simeq m_L M_E^{-1} m_R =(\kappa/\lambda) m_0 Z\cdot 
{\bf 1} \cdot Z$.

The mass spectra are essentially characterized by the parameter $b_f$.
The fermion masses $m_i^f$ versus $b_f$ are illustrated in Fig.~2.
At $b_f=0$, the charged lepton masses have been used as input values
for the parameters $z_i$.
Note that at $b_f=-1/3$, the third fermion mass takes a maximal 
value, which is independent of $\kappa/\lambda$.
Also note that at $b_f=-1/2$ and $b_f=-1$, two fermion masses 
degenerate.

%%%%%%%%%%%%%%%%%%%%%%%%%%%%%%%%%%%%%%%%%%%%%%%%%%%%%Fig.2
%\begin{figure}[htbp]
\begin{minipage}[tl]{10cm}
\epsfile{file=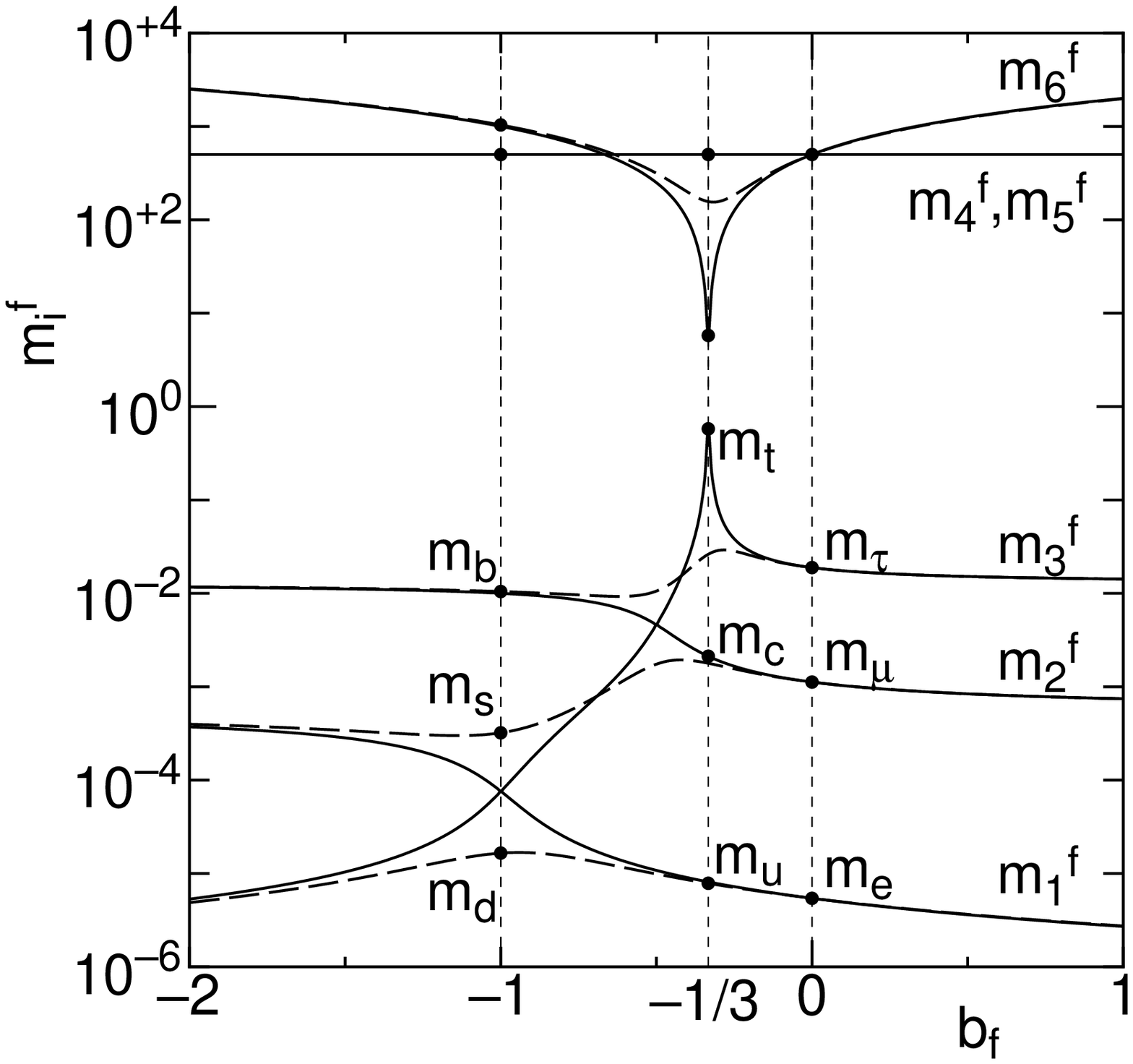,scale=0.45}
\end{minipage}
\begin{minipage}[tr]{5.5cm}
{\small Fig.~2. Masses $m_i$ ($i=1,2,\cdots,6$) versus $b_f$ for the 
case $\kappa=10$ and $\kappa/\lambda=0.02$.
The solid and broken lines represent the cases arg$b_f=0$ and 
arg$b_f=18^\circ$, respectively. The figure was quoted from Ref.~[13].
}
\end{minipage}
%\end{figure}
%%%%%%%%%%%%%%%%%%%%%%%%%%%%%%%%%%%%%%%%%%%%%%%%%%%%%%%

We take $b_u = -1/3$  for up-quark sector, 
because, at $b_u=-1/3$, we can obtain the maximal top-quark mass 
enhancement (see Fig.~2)
$$
m_t \simeq \frac{1}{\sqrt{3}} m_0 \ ,
\eqno(3.5)
$$
and  a successful relation
$$
 \frac{m_u}{m_c} \simeq \frac{3}{4}\frac{m_e}{m_\mu}  \ , 
\eqno(3.6)
$$
independently of the value of $\kappa/\lambda$.

The value of $\kappa/\lambda$ is determine from the observed ratio
$m_c/m_t$ as $\kappa/\lambda=0.0198$. 
Considering the successful relation 
$$
\frac{m_d m_s}{m_b^2}\simeq 4 \frac{m_e m_\mu}{m_\tau^2} \ ,
\eqno(3.7)
$$ 
for $b_d\simeq -1$, we seek for the best fit point of
$ b_d=- e^{i\beta_d}$ ($\beta_d^2 \ll 1$).
The observed ratio $m_d/m_s$ fixes the value $\beta_d$ 
as $\beta_d=18^\circ$.  
Then we can obtain the reasonable quark mass ratios [4],
not only $m_i^u/m_j^u$, $m_i^d/m_j^d$,  but also 
$m_i^u/m_j^d$:
$$
\begin{array}{lll}
m_u=0.000234\ {\rm GeV}, & m_c=0.610\ {\rm GeV}, &
m_t=0.181\ {\rm GeV}, \\
m_d=0.000475\ {\rm GeV}, & m_s=0.0923\ {\rm GeV}, &
m_b=3.01\ {\rm GeV}. \\
\end{array} 
\eqno(3.8)
$$
Here, we have taken 
$(m_0\kappa/\lambda)_q /(m_0\kappa/\lambda)_e=3.02$
in order to fit the observed quark mass values at 
$\mu=m_Z$ [14]
$$
\begin{array}{lll}
m_u=0.000233\ {\rm GeV}, & m_c=0.677\ {\rm GeV}, &
m_t=0.181\ {\rm GeV}, \\
{\ \ \ \ \ \ }_{-0.000045}^{+0.000042} &
{\ \ \ \ \ \ }_{-0.061}^{+0.056} & 
{\ \ \ \ \ \ }\pm 13 \\
m_d=0.000469\ {\rm GeV}, & m_s=0.0934\ {\rm GeV}, &
m_b=3.00\ {\rm GeV}. \\
{\ \ \ \ \ \ }_{-0.000066}^{+0.000060} &
{\ \ \ \ \ \ }_{-0.0130}^{+0.0118} &
{\ \ \ \ \ \ }\pm 0.11
\end{array} 
\eqno(3.9)
$$
We also obtain the reasonable values of the CKM
matrix parameters:
$$
\begin{array}{ll}
|V_{us}|=0.220 \ , \ \ \ &  |V_{cb}|=0.0598 \ , \\ 
|V_{ub}|=0.00330 \ , \ \ \ & |V_{td}|=0.0155 \ . 
\end{array} \eqno(3.10)
$$
(The value of $|V_{cb}|$ is somewhat larger than the observed value.
For the improvement of the numerical value, see Ref.~[13].)

%%%%%%%%%%%%%%%%%%%%%%%%%%%%%%
\vspace{.2in}

{\large\bf 4. Model A: a straightforward extension to the neutrinos}

\vglue.05in

The most straightforward extension of the model to the neutrinos is 
to consider the mass matrix of the neutrino sector is also given by
(1.1), so that the mass matrix $M_\nu$ for the conventional light 
neutrinos is given by (2.3), i.e.,
$$
M_\nu \simeq - m_L M_N^{-1} m_R = - \frac{\kappa}{\lambda_\nu} 
m_0 Z Y_\nu^{-1} Z\ .
\eqno(4.1)
$$
The smallness of the neutrino masses $m_\nu$ is given by 
assuming $\lambda_\nu \gg \lambda$ ($\lambda\equiv \lambda_e=
\lambda_u=\lambda_d$).

Our interest is in a large mixing solution.
As anticipated from Fig.~2, the large mixing solutions are
given at $b_f\simeq -1/2$ and $b_f\simeq -1$, at which
the mass degenerates $m_2^f\simeq m_3^f$ and 
$m_1^f\simeq m_2^f$ occur, respectively:

\noindent
[Case A] $b_\nu\simeq -1/2$: 
$$
m_1^\nu \simeq 2\frac{m_e}{m_\tau}\frac{\kappa}{\lambda_\nu}m_0 \ , 
\ \ \ m_2^\nu \simeq 
m_3^\nu \simeq \sqrt{\frac{m_\mu}{m_\tau}}
\frac{\kappa}{\lambda_\nu}m_0 \ , 
\eqno(4.2)
$$
\renewcommand{\arraystretch}{2}
$$
U\simeq \left(\begin{array}{ccc}
1 & \sqrt{\displaystyle\frac{m_e}{2m_\mu}} 
& \sqrt{\displaystyle\frac{m_e}{2m_\mu}} \\
-\sqrt{\displaystyle\frac{m_e}{m_\mu}} 
& \displaystyle\frac{1}{\sqrt{2}} & 
\mp \displaystyle\frac{1}{\sqrt{2}} \\
-\sqrt{\displaystyle\frac{m_e}{m_\tau}} 
& \pm \displaystyle\frac{1}{\sqrt{2}} & 
\displaystyle\frac{1}{\sqrt{2}} \\
\end{array} \right) \ . 
\eqno(4.3)
$$
\renewcommand{\arraystretch}{1}

\noindent
[Case B] $b_\nu\simeq -1$:
$$
m_1^\nu \simeq m_2^\nu \simeq \sqrt{\frac{m_e m_\mu}{m_\tau^2}}
\frac{\kappa}{\lambda_\nu}m_0 \ , \ \ \ m_3^\nu \simeq 
\frac{1}{2}\frac{\kappa}{\lambda_\nu}m_0 \ , 
\eqno(4.4)
$$
\renewcommand{\arraystretch}{2}
$$
U\simeq \left(\begin{array}{ccc}
\displaystyle\frac{1}{\sqrt{2}} & \mp \displaystyle\frac{1}{\sqrt{2}} 
& -\sqrt{\displaystyle\frac{m_e}{m_\tau}} \\
\pm \displaystyle\frac{1}{\sqrt{2}} 
& \displaystyle\frac{1}{\sqrt{2}} 
& -\sqrt{\displaystyle\frac{m_\mu}{m_\tau}} \\
\sqrt{\displaystyle\frac{m_\mu}{2m_\tau}} 
& \sqrt{\displaystyle\frac{m_\mu}{2m_\tau}} & 
1 \\
\end{array} \right) \ \ . 
\eqno(4.5)
$$
\renewcommand{\arraystretch}{1}

These cases are favorable to the large mixing picture 
suggested by the atmospheric neutrino data [11]. 
However, it is hard  to give the simultaneous 
explanation of the atmospheric and solar neutrino [12]
data, because the case $b_\nu\simeq -1/2$ 
($b_\nu\simeq -1$) can give $\sin^2 2\theta_{23}\simeq 1$
($\sin^2 2\theta_{12}\simeq 1$), while the case leads 
to $\Delta m^2_{atm}\equiv \Delta m^2_{32}
\ll \Delta m^2_{solar}\equiv \Delta m^2_{21}$ 
($\Delta m^2_{atm}\equiv \Delta m^2_{21}
\ll \Delta m^2_{solar}\equiv \Delta m^2_{32}$ ), 
where $\Delta m^2_{ij}=(m_i^\nu)^2-(m_j^\nu)^2$.
We must seek for another explanation for the solar 
neutrino data.

%%%%%%%%%%%%%%%%%%%%%%%%%%%%%%%%%%%%%%%%%%%%%%%%%
\vspace{.2in}

{\large\bf 5. Model B: a model without any new scale parameters}

\vglue.05in

The neutral lepton mass matrix which is sandwiched between 
$(\overline{\nu}_L, \overline{\nu}_R^c, \overline{N}_L, \overline{N}_R^c)$ 
and $(\nu_L^c, \nu_R, N_L^c, N_R)^T$, where $\nu_L^c \equiv (\nu_L)^c 
\equiv C \overline{\nu}_L^T$ and so on, is given by 
$$
M=\left(
\begin{array}{cccc}
0 & 0 & 0 & m_L \\
0 & 0 & m_R^T & 0 \\
0 & m_R & M_M & M_D \\
m_L^T & 0 & M_D^T & M_M 
\end{array} \right) \ , 
\eqno(5.1)
$$
where $M_D$ ($\equiv M_N$) and $M_M$ are Dirac and Majorana mass 
matrices of the neutral heavy fermions $N_i$.
The heavy fermions $F_i$ belong to $(1,1)$ of 
SU(2)$_L\times$U(2)$_R$.
Besides, the neutral heavy leptons $N_i$ do not have the U(1)-charge.
Therefore, it is likely that when the Dirac masses $(M_D)_{ij}$ 
are generated between $\overline{N}_{Li}$ and $N_{Rj}$, the Majorana 
masses $(M_M)_{ij}$ are also generated between $\overline{N}_{Li}$ and
$N_{Lj}^c$ ($\overline{N}_{Ri}^c$ and $N_{Rj}$) with the same structure 
at the same energy scale $\mu=\lambda m_0$. 
Hereafter, we assume that 
$$
M_M=M_D\equiv M_N\equiv \lambda m_0 Y_\nu \ .
\eqno(5.2)
$$
Then, we obtain the following twelve Majorana neutrinos [9]:
(i) three heavy Majorana neutrinos with masses of the order of 
$\lambda m_0$, whose mass matrix is approximately given by 
$$
M_{heavy}\simeq 2 M_N = 2\lambda m_0 Y_\nu \ , 
\eqno(5.3)
$$
(ii) three sets of almost degenerate two Majorana neutrinos 
(the pseudo-Dirac neutrino [6]) with masses of the order of 
$\kappa m_0$, whose mass matrix is approximately given by
\renewcommand{\arraystretch}{2}
$$
M_{PS-D} \simeq \left(
\begin{array}{cc}
-\frac{1}{4}m_R^T M_N^{-1} m_R & \displaystyle\frac{1}{\sqrt{2}} m_R^T \\
\displaystyle\frac{1}{\sqrt{2}} m_R & 0 \\
\end{array} \right) = \kappa m_0 \left(
\begin{array}{cc}
-\displaystyle\frac{\kappa}{4\lambda}Z Y_\nu^{-1} Z & 
\displaystyle\frac{1}{\sqrt{2}} Z \\
\displaystyle\frac{1}{\sqrt{2}} Z & 0 
\end{array} \right) \ ,
\eqno(5.4)
$$
\renewcommand{\arraystretch}{1}
and (iii) three light Majorana neutrinos with masses of the order of 
$(1/\lambda) m_0$,  whose mass matrix is approximately given by
$$
M_\nu \simeq - m_L M_N^{-1} m_L^T = 
-\frac{m_0}{\lambda} Z Y_\nu^{-1} Z \ .
\eqno(5.5)
$$

The neutrinos which are described by the mass matrix (5.5) 
consist of almost left-handed neutrinos $\nu_{Li}$. 
Therefore, as far as the conventional light neutrinos are 
concerned, the model B is identical with the model A by 
substituting $1/\lambda$ for $\kappa/\lambda_\nu$ in
the model A.
The numerical results have been given in Ref.~[9] in detail.
For example, for the case $b_\nu=-(1/2)e^{i\beta_\nu}$ 
($\beta_\nu=0.12^\circ$), we obtain 
$$
m_1^\nu=0.00164 \ {\rm eV}\ , \ \ \ \ 
m_2^\nu=0.695 \ {\rm eV}\ , \ \ \ \ 
m_3^\nu=0.707 \ {\rm eV}\ , 
\eqno(5.6)
$$
$$
\Delta m^2_{32}=0.016 \ {\rm eV}^2 \ , \ \ \ \ 
\Delta m^2_{21}=0.483 \ {\rm eV}^2 \ ,
\eqno(5.7)
$$
$$
\sin^2 2\theta_{atm}\equiv -4 {\rm Re} (U_{\mu 2}
U_{\tau 2}^* U_{\mu 3}^* U_{\tau 3}) = 0.995 \ ,
\eqno(5.8)
$$
$$ 
\sin^2 2 \theta_{LSND} \equiv 4 |U_{e1}|^2 |U_{\mu 1}|^2
= 0.0191 \ , 
\eqno(5.9)
$$
$$
\langle m_\nu \rangle \equiv \left| m_1^\nu U_{e1}^2+m_2^\nu U_{e2}^2
+m_3^\nu U_{e3}^2 \right| = 0.00267\ {\rm eV} \ .
\eqno(5.10)
$$ 
$$
\begin{array}{l}
\Lambda_L=m_0=3.12\times 10^2 \ {\rm GeV}\ , \\ 
\Lambda_R=\kappa m_0=1.90\times 10^{11} \ {\rm GeV}\ , \\ 
\Lambda_S=\lambda m_0=3.20\times 10^{13} \ {\rm GeV}\ . 
\end{array}
\eqno(5.11)
$$ 
Here, the numerical result (5.11) has been obtained from 
the input value $\Delta m^2_{atm}\equiv \Delta m^2_{32}=
0.016$ eV$^2$ and the relations 
$$
m_t\simeq \frac{1}{\sqrt{3}}  m_0 \simeq 180\  {\rm GeV} \ \ 
\ \  ({\rm at}\ \mu=m_Z) \ , 
\eqno(5.12)
$$
and  
$$
(\kappa/\lambda) m_0 = m_\tau + m_\mu +m_e =1.850 \ {\rm GeV} \ ,
\ \  ({\rm at}\ \mu=m_Z) \ .
\eqno(5.13)
$$
The results (5.7) - (5.9) can successfully explain the atmospheric 
neutrino data [11] and the neutrino oscillation 
$(\overline{\nu}_\mu\rightarrow\overline{\nu}_e)$ experiment
by the liquid scintillator neutrino detector (LSND) [15] 
at Los Alamos.
However, the model B fails to explain the solar neutrino data
straightforwardly, as well as the model A.

%%%%%%%%%%%%%%%%%%%%%%%%%%%%%%%%%%%%%%%%%%%%%%%%%
\vspace{.2in}

{\large\bf 6. Model C: a model with light pseudo-Dirac neutrinos}

\vglue.05in

In the present model, there is no distinction between
$N_L$ and $N_R$, because both fields are 
SU(2)$_L\times$SU(2)$_R$ singlets and do not have 
U(1) charges.
Therefore, if $\nu_L$ ($\nu_R$) acquire masses $m_L$ ($m_R$)
together with the partners $N_R$ ($N_L$), they may also 
acquire masses $m'_L$ ($m'_R$) together with the partners 
$N_L^c$ ($N_R^c$).
Then, the mass matrix for the neutrino sector is given by
$$
M=\left(
\begin{array}{cccc}
0 & 0 & m'_L & m_L \\
0 & 0 & m_R^T & m^{\prime T}_R \\
m^{\prime T}_L & m_R & M_M & M_D \\
m_L^T & m'_R & M_D^T & M_M 
\end{array} \right) \ . 
\eqno(6.1)
$$
By rotating the fields 
$(\nu_L^c, \nu_R, N_L^c, N_R)$ by 
$$
R_{34}=\left(
\begin{array}{cccc}
1 & 0 & 0 & 0 \\
0 & 1 & 0 & 0 \\
0 & 0 & \frac{1}{\sqrt{2}} & -\frac{1}{\sqrt{2}} \\
0 & 0 & \frac{1}{\sqrt{2}} & \frac{1}{\sqrt{2}} 
\end{array}\right) \ , 
\eqno(6.2)
$$
the $12\times 12$ mass matrix (6.1) becomes 
$$
M'= R_{34} M R_{34}^T = \frac{1}{\sqrt{2}} \left(
\begin{array}{cccc}
0 & 0 & m'_L-m_L & m'_L+m_L \\
0 & 0 & (mR-m'_R)^T & (m_R+m'_R)^T \\
(m'_L-m_L)^T & m_R-m'_R & \sqrt{2}(M_M-M_D) & 0 \\
(m'_L+m_L)^T & m_R+m'_R & 0 & \sqrt{2}(M_M+M_D) 
\end{array} \right) \ ,
\eqno(6.3)
$$
where we have used $M_D^T=M_D$.

In the $N_L\leftrightarrow N_R^c$ symmetric limit, i.e., 
in the limit of  $m'_L=m_L$,  $m'_R=m_R$ and 
$M_M=M_D\equiv M_N$, 
the mass matrices for the neutrinos $N_1$, $N_2$ and
$\nu_R$ are given by
$$
M(N_1)=0 \ ,
\eqno(6.4)
$$
$$
M(N_2)\simeq 2 M_N \ ,
\eqno(6.5)
$$
$$
M(\nu_R) \simeq - m_R^T M_N^{-1} m_R \ ,
\eqno(6.6)
$$
where $N_1$ and $N_2$ are defined by
$$
N_{1i}=\frac{1}{\sqrt{2}}(\nu_{Li} - N_{Ri}^c) \ , \ \ \ 
N_{2i}=\frac{1}{\sqrt{2}}(\nu_{Li} + N_{Ri}^c) \ .
\eqno(6.7)
$$
Furthermore, for the model with the relation $m_R\propto m_L$
such as a model given in Sec.~3, we obtain
$$
M(\nu_L) =0 \ .
\eqno(6.8)
$$
Only when we assume a sizable difference between 
$m'_L$ and $m_L$ (and also between 
$m'_R$ and $m_R$), we can obtain visible neutrino
masses $m(\nu_L)\neq 0$ and $m(N_1)\neq 0$.
A suitable choice of the differences 
$m'_L-m_L$ and $m'_R-m_R$ can lead to 
three sets of the pseudo-Dirac neutrinos
$(\nu^{ps}_{i+}, \nu^{ps}_{i-})$ ($i=e, \mu, \tau$)
which are large mixing states between $\nu_{Li}$ 
and $N_{1i}$, i.e.,
$$
\nu^{ps}_{i\pm} \simeq \frac{1}{\sqrt{2}}
(\nu_{Li}\pm N_{1i}) \ ,
\eqno(6.9)
$$
and whose masses are almost degenerate.
The ps-Dirac neutrinos $\nu^{ps}_{\mu \pm}$ are 
favorable to the explanation of the large mixing 
suggested from the atmospheric neutrino date [11] 
and the ps-Dirac neutrinos $\nu^{ps}_{e \pm}$ are 
favorable to that of the large-angle solution 
of the Mikheyev-Smirnov-Wolfenstein (MSW) 
solutions [16] for the solar neutrino data [12].
(A suitable choice can also give that the lightest two 
neutrino states are favorable to the small-angle
solution of the MSW solutions.) 
The numerical results from the model C will be given 
elsewhere in the collaboration with Fusaoka [17].

If the scenario C is true, we will not able to observe
a large-angle mixing in appearance experiments such as 
$\nu_\mu \rightarrow\nu_\tau$ oscillation experiments 
by CHORUS [18] and NOMAD [19].
The large-angle mixing will be observed only in
the disappearance experiments, because the 
mixing partners of $\nu_\mu$ and $\nu_e$ are
SU(2)$_L\times$SU(2)$_R$ singlet neutrinos
$N_{1\mu}$ and $N_{1e}$, respectively.

However, for example, when we take 
$m'_L=(1-\varepsilon_L)m_L$ and 
$m'_R=(1-\varepsilon_R)m_R$ [$\varepsilon=$
diag$(\varepsilon_1,\varepsilon_2,\varepsilon_3)$],
the these $\varepsilon$ are must be of the order of
$10^{-11}$. 
Where such the scale of $\varepsilon\sim 10^{-11}$ 
comes from is open question in the scenario C.

%%%%%%%%%%%%%%%%%%%%%%%%%%%%%%%%%%%%%%%%%%%%%%%%%
\vspace{.2in}

{\large\bf 7. Concluding remarks}

\vglue.05in

In conclusion, we have discussed possible three 
neutrino-mass-generation scenarios within the framework of the
universal seesaw mechanism, especially, on the basis
of the ``democratic seesaw mass matrix model", which
can answer the question why only top quark $t$ acquires
the mass of the order of the electroweak scale 
$\Lambda_L=O(m_L)$, and can give reasonable quark 
masses and mixings in terms of the charged lepton 
masses.

The scenario A is a trivial extension of the model for
quarks, and it is not so attractive, because the 
smallness of the neutrino masses is explained by
introducing an additional energy scale parameter
$\Lambda_S^\nu$ ($\Lambda_S^\nu\gg \Lambda_S
\equiv \Lambda_S^e=\Lambda_S^u=\Lambda_S^d$),
by hand.
In contrast with the scenario A, the scenario B
is a model without a new energy scale such as 
$\Lambda_S^\nu$.
The smallness of the neutrino masses is explained
by the suppression factor $1/\kappa\equiv \Lambda_R/
\Lambda_S$.
The choice of the parameter value $b_\nu\simeq -1/2$
can give reasonable values of $\Delta m^2_{atm}=
\Delta m^2_{32}\sim 10^{-2}$ eV$^2$ and 
$\sin^2 2\theta_{23}\simeq 1$, but it fails to
explain the solar neutrino data.

The scenario C is the most attractive model from 
the phenomenological point of view.
The model leads to three sets of the pseudo-Dirac
neutrinos $(\nu^{ps}_{i+},\nu^{ps}_{i-})$ 
($i=e,\mu,\tau$), which are large mixing states
between $\nu_{Li}$ and $N_{1i}\equiv(N_{Li}-
N_{Ri}^c)/\sqrt{2}$.
If the scenario C is true, we will not be able 
to observe a large-angle mixing in the appearance
oscillation experiments such as $\nu_\mu\rightarrow
\nu_\tau$ oscillation experiments by CHORUS and
NOMAD.
The large-angle mixing will be observed only in
the disappearance experiments, because the 
mixing partners of $\nu_\mu$ and $\nu_e$ are
SU(2)$_L\times$SU(2)$_R$ singlet neutrinos
$N_{1\mu}$ and $N_{1e}$, respectively.
However, in the scenario C, the smallness of
the neutrino masses must be explained by a 
small breaking of the $N_L\leftrightarrow 
N_R^c$ symmetry, by hand.
This is our future task.

At present, it seems to be hard to explain
all of the neutrino data within the framework
of the conventional light three family 
neutrinos.
Since the scenario C leads to three sets of 
the light ps-Dirac neutrinos (i.e., six 
almost massless neutrinos), the model can 
give us abundant phenomenological predictions.
It will be worth while studying the scenario C 
furthermore from the phenomenological point of 
view as well as from the theoretical point of view.

%%%%%%%%%%%%%%%%%%%%%%%%%%%%%%%%%%%%%%%%%%%
\vspace*{.2in}
\centerline{\large\bf Acknowledgements}

\vglue.05in

The author would like to thank Prof.~H.~Fusaoka for 
his enjoyable and  powerful collaboration.
This work was supported by the Grand-in-Aid for Scientific
Research, Ministry of Education, Science and Culture,
Japan (No.~08640386).

%\newpage
%%%%%%%%%%%%%%%%
\vglue.2in
\newcounter{0000}
\centerline{\large\bf References}
\begin{list}
{[~\arabic{0000}~]}{\usecounter{0000}
\labelwidth=0.8cm\labelsep=.1cm\setlength{\leftmargin=0.7cm}
{\rightmargin=.2cm}}
\item Z.~G.~Berezhiani, Phys.~Lett.~{\bf 129B}, 99 (1983);
Phys.~Lett.~{\bf 150B}, 177 (1985);
D.~Chang and R.~N.~Mohapatra, Phys.~Rev.~Lett.~{\bf 58},1600 (1987); 
A.~Davidson and K.~C.~Wali, Phys.~Rev.~Lett.~{\bf 59}, 393 (1987);
S.~Rajpoot, Mod.~Phys.~Lett. {\bf A2}, 307 (1987); 
Phys.~Lett.~{\bf 191B}, 122 (1987); Phys.~Rev.~{\bf D36}, 1479 (1987);
K.~B.~Babu and R.~N.~Mohapatra, Phys.~Rev.~Lett.~{\bf 62}, 1079 (1989); 
Phys.~Rev. {\bf D41}, 1286 (1990); 
S.~Ranfone, Phys.~Rev.~{\bf D42}, 3819 (1990); 
A.~Davidson, S.~Ranfone and K.~C.~Wali, 
Phys.~Rev.~{\bf D41}, 208 (1990); 
I.~Sogami and T.~Shinohara, Prog.~Theor.~Phys.~{\bf 66}, 1031 (1991);
Phys.~Rev. {\bf D47}, 2905 (1993); 
Z.~G.~Berezhiani and R.~Rattazzi, Phys.~Lett.~{\bf B279}, 124 (1992);
P.~Cho, Phys.~Rev. {\bf D48}, 5331 (1994); 
A.~Davidson, L.~Michel, M.~L,~Sage and  K.~C.~Wali, 
Phys.~Rev.~{\bf D49}, 1378 (1994); 
W.~A.~Ponce, A.~Zepeda and R.~G.~Lozano, 
Phys.~Rev.~{\bf D49}, 4954 (1994).
\item M.~Gell-Mann, P.~Rammond and R.~Slansky, in {\it Supergravity}, 
edited by P.~van Nieuwenhuizen and D.~Z.~Freedman (North-Holland, 
1979); 
T.~Yanagida, in {\it Proc.~Workshop of the Unified Theory and 
Baryon Number in the Universe}, edited by A.~Sawada and A.~Sugamoto 
(KEK, 1979); 
R.~Mohapatra and G.~Senjanovic, Phys.~Rev.~Lett.~{\bf 44}, 912 (1980).
\item CDF Collaboration, F.~Abe {\it et al.}, Phys.~Rev.~Lett. 
{\bf 73}, 225 (1994).
\item Y.~Koide and H.~Fusaoka, Z.~Phys. {\bf C71}, 459 (1996). 
\item T.~Morozumi, T.~Satou, M.~N.~Rebelo and M.~Tanimoto, Phys.~Lett. 
{\bf B410}, 233 (1997).
\item L.~Wolfenstein, Nucl.~Phys. {\bf B185}, 147 (1981); 
S.~T.~Petcov, Phys.~Lett. {\bf 110B}, 245 (1982);
C.~N.~Leung and S.~T.~Petcov, Nucl.~Phys. {\bf 125B}, 461 (1983); 
M.~Doi, M.~Kenmoku, T.~Kotani, H.~Nishiura and E.~Takasugi, 
Prog.~Theor.~Phys. {\bf 70}, 1331 (1983);
J.~W.~F.~Valle, Phys.~Rev. {\bf D27}, 1672 (1983);
J.~W.~F.~Valle and M.~Singer, Phys.~Rev. {\bf D28}, 540 (1983);
D.~Wyler and L.~Wolfenstein, Nucl.~Phys. {\bf B218}, 205 (1983).
\item N.~Cabibbo, Phys.~Rev.~Lett.~{\bf 10}, 531 (1963); 
M.~Kobayashi and T.~Maskawa, Prog.~Theor.~Phys.~{\bf 49}, 652 (1973).
\item Z.~G.~Berezhiani, in Ref.[2]; 
A.~Davidson and K.~C.~Wali, in Ref.[2];
S.~Rajpoot, in Ref.[4];
A.~Davidson, S.~Ranfone and K.~C.~Wali, in Ref.[2];
W.~A.~Ponce, A.~Zepeda and R.~G.~Lozano, in Ref.[2].
\item Y.~Koide, Phys.~Rev. {\bf D57}, 5836 (1998). 
\item Y.~Koide, Phys.~Rev. {\bf D56}, 2656 (1997).
\item Y.~Fukuda {\it et al.}, Phys.~Lett. {\bf B335}, 237 (1994);
also see, Soudan-2 Collaboration, M.~Goodman {\it at al}., in 
{\it Neutrino 94}, Proceedings of the 16th International 
Conference on Neutrino Physics and Astrophysics, Eilat, Israel, 
edited by A.~Dar {\it at al}. [Nucl.~Phys. {\bf B} (Proc.~Suppl.) 
{\bf 38}, 337 (1995)]; IBM Collaboration, D.~Casper {\it et al}.,
Phys.~Rev.~Lett. {\bf 66}, 2561 (1989); R.~Becker-Szendy 
{\it et al}., Phys.~Rev. {\bf D 46}, 3729 (1989). 
\item GALLEX collaboration, P.~Anselmann {\it et al.}, 
Phys.~Lett. {\bf B327}, 377 (1994); {\bf B357}, 237 (1995); 
SAGE collaboration, J.~N.~Abdurashitov {\it et al.}, {\it ibid.} 
{\bf B328}, 234 (1994); also see N.~Hata and P.~Langacker, 
Phys.~Rev. {\bf D50}, 632 (1994); {\bf D52}, 420 (1995).
\item Y.~Koide and H.~Fusaoka, 
Prog.~Theor.~Phys. {\bf 97}, 459 (1997).
\item H.~Fusaoka and Y.~Koide, Phys.~Rev. {\bf D57}, 3986 (1998).
\item C.~Athanassopoulos {\it et al.}, Phys.~Rev.~Lett. {\bf 75}, 2650 
(1995);  Phys.~Rev.~Lett. {\bf 77}, 3082 (1996); nucl-ex/9706006 (1997).
\item S.~P.~Mikheyev and A.~Yu.~Smirnov, Yad.~Fiz. {\bf 42}, 1441 (1985); 
[Sov.~J.~Nucl.~Phys. {\bf 42}, 913 (1985)]; 
Prog.~Part.~Nucl.~Phys. {\bf 23}, 41 (1989); 
L.~Wolfenstein, Phys.~Rev. {\bf D17}, 2369 (1978); {\bf D20}, 2634 (1979);
T.~K.~Kuo and J.~Pantaleon, Rev.~Mod.~Phys. {\bf 61}, 937 (1989). 
Also see, A.~Yu.~Smirnov, D.~N.~Spergel and J.~N.~Bahcall, 
Phys.~Rev. {\bf D49}, 1389 (1994).
\item Y.~Koide and H.~Fusaoka, in preparation, US-98-06.
\item M.~Baldo-Ceolin, in {\it TAUP 93}, Proceedings of the Third
International Workshop on Theoretical and Phenomenological Aspects
of Underground Physics, Assergi, Italy, edited by C.~Arpesella 
{\it et al}. [Nucl.~Phys. Proc.~Suppl. {\bf 35}, 450 (1994)]; 
K.~Winter, in {\it Neutrino '94} (Ref.~[11]), p.~211. 
\item L.~DiLella, in {\it Neutrino 92}, Proceedings of the 
XVth International Conference on Neutrino Physics and Astrophysics, 
Granada, Spain, edited by A.~Morales [Nucl.~Phys. Proc.~Suppl. 
{\bf 31}, 319 (1993)].
%
%\item Y.~Koide and M.~Tanimoto, Z.~Phys. {\bf C72}, 333 (1996);
%A.~Davidson, T.~Schwartz and R.~R.~Volkas, hep-ph/9802235, (1998).
%
%\item Y.~Koide, Mod.~Phys.~Lett. {\bf A36}, 2849 (1996).
%
%
%\item H.~Harari, H.~Haut and J.~Weyers, Phys.~Lett. {\bf 78B}, 459 (1978). 
%
\end{list}

\end{document}